\documentclass[pdflatex,sn-nature]{sn-jnl}
\usepackage{graphicx}%
\usepackage{multirow}%
\usepackage{amsmath,amssymb,amsfonts}%
\usepackage{amsthm}%
\usepackage{mathrsfs}%
\usepackage[title]{appendix}%
\usepackage{xcolor}%
\usepackage{textcomp}%
\usepackage{manyfoot}%
\usepackage{booktabs}%
\usepackage{algorithm}%
\usepackage{algorithmicx}%
\usepackage{algpseudocode}%
\usepackage{listings}%

\begin{document}

\title[Article Title]{Phonon-enhanced strain sensitivity of quantum dots in two-dimensional semiconductors}


\author[1]{\fnm{Sumitra} \sur{Shit}}

\author[1]{\fnm{Yunus} \sur{Waheed}}

\author[1]{\fnm{Jithin Thoppil} \sur{Surendran}}

\author[1]{\fnm{Indrajeet Dhananjay} \sur{Prasad}}

\author[2]{\fnm{Kenji} \sur{Watanabe}}

\author[3]{\fnm{Takashi} \sur{Taniguchi}}

\author*[1]{\fnm{Santosh} \sur{Kumar}}\email{skumar@iitgoa.ac.in}

\affil*[1]{\orgdiv{School of Physical Sciences}, \orgname{Indian Institute of Technology Goa}, \orgaddress{\city{Ponda}, \postcode{403401}, \state{Goa}, \country{India}}}

\affil[2]{\orgdiv{Research Center for Electronic and Optical Materials}, \orgname{National Institute for Materials Science}, \orgaddress{\street{1-1 Namiki}, \city{Tsukuba}, \postcode{305-0044}, \country{Japan}}}

\affil[3]{\orgdiv{Research Center for Materials Nanoarchitectonics}, \orgname{National Institute for Materials Science}, \orgaddress{\street{1-1 Namiki}, \city{Tsukuba}, \postcode{305-0044}, \country{Japan}}}

\abstract{Two-dimensional semiconductors have attracted considerable interest for integration into emerging quantum photonic networks. Strain engineering of monolayer transition-metal dichalcogenides (ML-TMDs) enables the tuning of light-matter interactions and associated optoelectronic properties, and generates new functionalities, including the formation of quantum dots (QDs). Here, we combine spatially resolved micro-photoluminescence ($\mu$-PL) spectroscopy from cryogenic (4--94\,K) to room temperature with micro-Raman spectroscopy at room temperature to investigate the strain-dependent emission energies of thousands of individual QDs in ML-WS$_2$ and ML-WSe$_2$, integrated across multiple heterostructures and a piezoelectric device. Compared with delocalized excitons, QDs in both materials exhibit enhanced strain sensitivities of their emission energies\textemdash approximately fourfold in WS$_2$ and twofold in WSe$_2$\textemdash leading to pronounced broadening of the ensemble emission linewidth. Temperature-dependent $\mu$-PL spectroscopy combined with dynamic strain tuning experiments further reveal that the enhanced strain sensitivity of individual QDs originates from strengthened interactions with low-energy phonons induced by quantum confinement. Our results demonstrate a versatile strain-engineering approach with potential for spectral matching across solid-state, atomic, and hybrid quantum photonic networks, and provide new insights into phonon-QD interactions in two-dimensional semiconductors.}


\maketitle

\section*{Main}
Quantum dots (QDs) in two-dimensional (2D) semiconductors have emerged as a promising solid-state nanophotonic platform for next-generation quantum photonic technologies\cite{aharonovich2016solid,montblanch2023layered}. Remarkable functionalities\textemdash including high\textendash purity single\textendash photon emission\cite{kumar2016resonant,piccinini2025high}, entangled photon\textendash pair generation\cite{he2016biexciton,chen2019entanglement}, and Coulomb blockade of individual electrons and holes\cite{brotons2019coulomb}—have been demonstrated in QDs hosted by atomically thin transition-metal dichalcogenides (TMDs), a widely studied 2D semiconductor platform. QDs in monolayer (ML) and few-layer TMDs typically exhibit either comb-like emission spectra or a small number of spectrally isolated lines with widely varying wavelengths. This intrinsic spectral variability hinders the realization of multiple QDs with identical optical properties, which are essential for next-generation scalable quantum networks. Consequently, post-fabrication tuning strategies are required to achieve wavelength-tunable QDs in TMDs for developing hybrid quantum photonic networks that integrate TMD QDs with established solid-state III-V QDs and atomic systems.

To address this spectral inhomogeneity, various tuning approaches have been explored. These include, but are not limited to, temperature tuning\cite{reithmaier2004strong}, strain tuning\cite{seidl2006effect}, laser annealing\cite{yu2025dynamic}, vertical electric field tuning\cite{bennett2010electric}, and magnetic field tuning\cite{akopian2010tuning}. As electric\cite{ramsay2008fast,Godden2012Coherent} and magnetic\cite{Shopia2007Theory} fields play crucial roles in initialization, control and manipulation of electron or hole spins in QDs, strain tuning provides a complementary and effective route for wavelength control. Strain modifies the electronic band structure of TMDs and affects the light-matter interaction, influencing the exciton localization\cite{gelly2022probing} and its emission wavelength\cite{aslan2018strain,waheed2025large}.

Strain not only tunes the exciton emission wavelength but also reshapes the phonon landscape. The role of phonon interactions in determining the functionalities of QDs across a wide range of material platforms is well documented. For example, two-phonon processes combined with other interactions\cite{Trif09} have been shown to account for long spin relaxation times\textemdash approaching 1 ms\textemdash of heavy holes confined in silicon QDs\cite{Heiss07, Hanson07} as well as in III\textendash V QDs\cite{Gerardot08, michler2000quantum,huo2014light,kako2006gallium}. More recently, chiral phonons in ML-WSe$_2$ have been demonstrated to provide a pathway for the generation of entangled photon pairs\cite{chen2019entanglement}. These advances suggest that the range of functionalities of QDs in ML-TMDs can be further extended through both strain engineering and phonon engineering, including the activation of chiral phonons\cite{pan2024strain}.

Here, we investigate the pronounced strain sensitivity of QD emission energies in ML-TMDs using low-temperature (4\,K), room-temperature (296\,K), and temperature-dependent spatially resolved micro-photoluminescence ($\mu$-PL) spectroscopy. A large number of individual QDs integrated into distinct van der Waals heterostructures, as well as a piezoelectric device, were investigated. QDs in ML-WS$_2$ and ML-WSe$_2$ are defined and tuned simultaneously by localized strain pockets created by placing ML-TMDs flakes on spherical SiO$_2$ nanoparticles (SNPs), e-beam-deposition-engineered SiO$_x$ nanoparticles (ENPs) and nanodroplets (NDs), or by unintentional wrinkles. We explored strain ranges of -0.10\textendash 0.75\% for ML-WS$_2$ QDs and 0.05\textendash 0.20\% for ML-WSe$_2$ QDs and show that, compared with delocalized excitons $\left(\text{2D-X}^0\right)$ in these ML-TMDs, the emission energies of the QDs exhibit significantly larger strain-induced shift rates. This enhanced sensitivity leads to a pronounced broadening of the ensemble emission linewidth (full width at half maximum, FWHM). Dynamic strain tuning and temperature-dependent $\mu$-PL measurements further reveal that this pronounced strain sensitivity\textemdash and the resulting broadening\textemdash are attributed to a quantum-confinement-induced enhancement of exciton-(low-energy acoustic) phonon interactions.
\begin{figure*}[htb]
\centering
\includegraphics[width=0.95\textwidth]{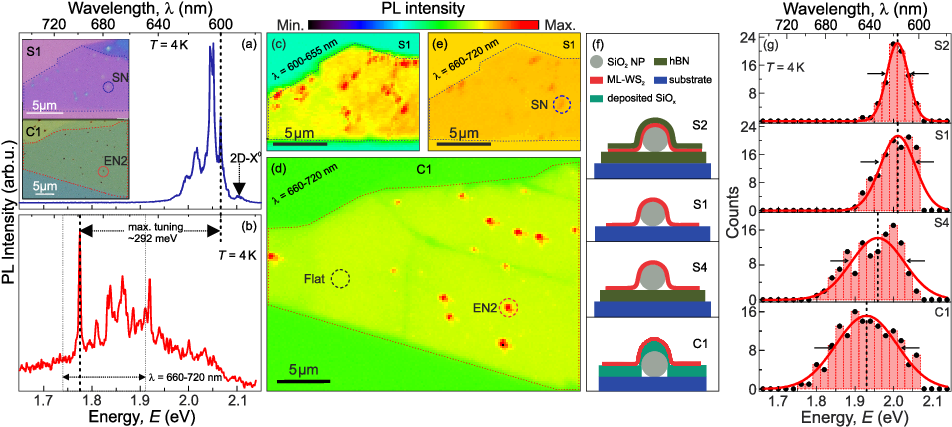}
\caption{\textbf{Redshifted quantum dots (QDs) and ensemble broadening in ML-WS$_2$.\,}Representative $\mu$-PL spectrum of QDs at (a) a spherical nanoparticle (SNP) location in sample\,S1 and (b) a shape-engineered nanoparticle (ENP) location in sample\,C1. Vertical dotted lines indicate the highest and lowest QD emission energies observed across the PL spectra in (a) and (b), corresponding to an energy span of approximately 292 meV. Inset in (a): Optical micrographs of samples\,S1 (top) and C1 (bottom), with ML-WS$_2$ flakes outlined by dotted lines. (c-e) Spatial maps of $\mu$-PL peak intensity of QDs in ML-WS$_2$ at low temperature ($T\,=\,4\,$K) from samples (c)\,S1 in the 600\textendash 655\,nm range, (d)\,C1 in the 660\textendash 720\,nm range, and (e)\,S1 in the 660\textendash 720\,nm range, revealing that highly redshifted QDs are observed only at ENP locations. (f) Schematics of four different sample structures containing SNPs or ENPs as nanostressors. (g) Histograms of QD emission energies from the four samples shown in (f), illustrating ensemble emission spanning the 1.81\textendash 2.08 eV range (bin size: 20\,meV). Solid curves are the Gaussian fits, vertical dotted lines mark the ensemble peak energies, and arrows indicate the ensemble full width at half maximum linewidths.}
\label{Fig:fig1}
\end{figure*}

\section*{Results}
\noindent\textbf{Highly redshifted QD in ML-WS$_2$ due to a shape engineered nanostressor:\,}Our investigation begins with 4\,K $\mu$-PL spectra of ML-WS$_2$ acquired at individual SNP (SN) and ENP (EN2) locations (Figs.\,\ref{Fig:fig1}a\textendash b; see optical micrographs in the inset of Fig.\,\ref{Fig:fig1}a marking SN and EN locations). The sharp emission lines observed in these PL spectra arise from localized excitons and are indicative of emission from a few QDs in ML-WS$_2$\cite{palacios2017large,surendran2024nanoparticle}. Notably, the PL spectrum taken at the EN2 location shows emission extending beyond 700\,nm, including a sharp emission line at 1.775\,eV. This emission is strongly redshifted compared with a representative QD emission line at 2.067\,eV at the SN location. These observations demonstrate that NP shape-engineering enables tuning of QD emission energies in ML-WS$_2$ over a maximum range of $\approx$\,292\,meV.

We next examine spatially resolved PL emission to show that these highly redshifted QDs are present exclusively at ENP locations and are absent at SNP locations. Recent studies report that QD emission in ML-WS$_2$ occurs in the 605\textendash 655\,nm wavelength range \cite{palacios2017large,surendran2024nanoparticle}. Consistent with this, we observe QD emission in ML-WS$_2$ at SNP locations in sample\,S1 through bright spots in a $\mu$-PL map of peak intensities within the same spectral range (600\textendash 655\,nm; Fig.\,\ref{Fig:fig1}c). In contrast, PL intensity maps acquired at longer wavelengths (660\textendash 720\,nm) reveal a pronounced dependence on nanoparticle (NP) type. Highly red-shifted QD emission is observed exclusively at ENP locations (Figure\,\ref{Fig:fig1}d for sample C1), while being absent at SNP locations (Figure\,\ref{Fig:fig1}e for sample S1).\\

\begin{figure}[htb]
\centering
\includegraphics[width=8.5cm]{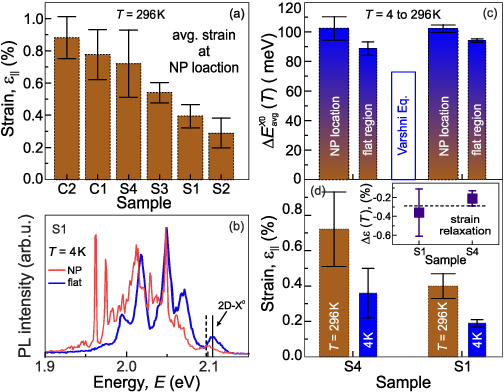}
\caption{\textbf{Thermoelastic strain relaxation at the NP location upon cooling from room temperature (RT) to 4\,K.\,}(a) Bar plot of the average local in-plane strain\,$\left(\epsilon_\parallel\right)$ in ML-WS$_2$ at the NP location for six samples, showing a biaxial strain range of 0.30\textendash 0.90\% at RT. (b) 4\,K $\mu$-PL spectra of ML-WS$_2$ taken at the NP (thin curve) and flat (thick curve) locations in sample\,S1. The vertical dotted (solid) line marks the X$^0$ emission energy. (c) Bar plot of average X$^0$ energy shifts at NP (wide bars) and flat (narrow bars) locations for samples S1 and S4 upon cooling to 4\,K, compared with the expected energy shift (open bar) according to Varshni equation. (d) Bar plot comparing $\epsilon_\parallel$ at RT and 4\,K at the NP location for samples\,S1 and S4. Inset: Thermoelastic strain relaxation upon cooling from RT to 4\,K for these samples. Error bars in (a), (c), and (d) are described in the Methods section.}
\label{Fig:fig2}
\end{figure}
\noindent\textbf{Strain-induced redshift and linewidth broadening of QD ensembles in ML-WS\textsubscript{2}:\,}To understand the origin of the highly red-shifted QD emission, we analysed sharp emission lines of QDs from six samples incorporating either ENPs or SNPs as strain inducers, including samples C1 and S1. PL intensity maps of four additional samples were also acquired; all show multiple bright NP locations, each exhibiting a few emission lines (Supplementary Figs.\,S1\textendash S2). The statistical distribution of QD emission energies for samples\,S1, S2, S4, and C1 is summarised in Fig.\,\ref{Fig:fig1}g through histogram plots, while data for samples S3 and C2 are given in Supplementary Figs.\,S7b\textendash c. The bin counts (closed circles) were fitted with a Gaussian function to extract ensembles' peak energies and FWHM linewidths. Ensemble-like QD emissions in all these histograms show systematic redshifts and linewidth broadenings across different samples, suggesting strong variations in local strain at NP locations in these samples.

While different layer structures\cite{martin2020encapsulation}, incorporating ML-TMDs, have been reported to improve the optical quality of excitons, in our study they play a crucial role in producing samples with varying average local strain at the NP locations, as shown in the column plot in Fig.\,\ref{Fig:fig2}a. Local strains at RT were estimated using $\mu$-PL emission acquired at all NP locations in each sample. $\mu$-Raman spectroscopy on sample\,C1 confirms the PL-derived strain within a 0.18\% standard deviation; therefore, PL analysis was used for all samples (Supplementary Sec. II, and Tab.\,S1). The various bright spots in the PL spatial maps of samples S1 and C1 (Figs.\ref{Fig:fig1}c\textendash d) originate from localised strain in the ranges of 0.16\,-\,0.68\% and 0.30\,-\,0.96\% (Supplementary Fig.\,S6, and Fig.\,S4), respectively.\\

\begin{figure}[htb]
\centering
\includegraphics[width=8.0cm]{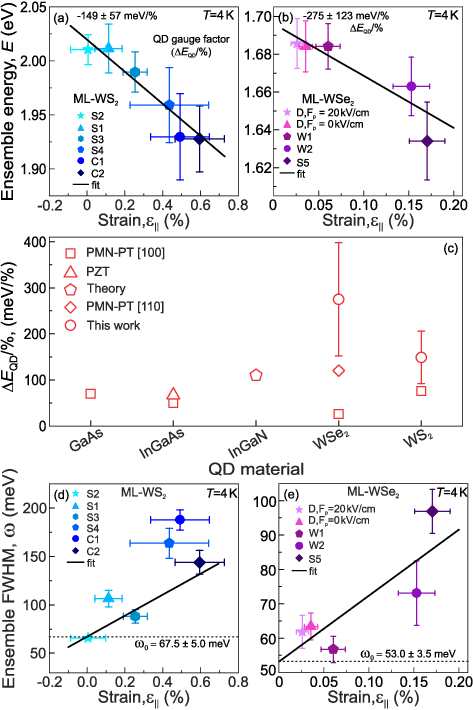}
\caption{\textbf{Correlations of emission energy and linewidth of QD ensemble with localized strain at 4\,K.\,}Strain-dependent emission energies of QD ensembles in (a) ML-WS$_2$ from six different samples,  and (b) ML-WSe$_2$ from four different samples. Different symbols represent measurements from each sample; the solid line is a linear fit. (c) Comparison of gauge factors (strain sensitivity of emission energy) across multiple QD platforms; symbol shapes represent the tuning techniques. Strain-dependent linewidth of QD ensemble in (d) ML-WS$_2$ across six samples, and (e) ML-WSe$_2$ across four samples, showing significant strain-induced QD ensemble broadening in both materials. Closed symbols represent measured data; solid lines are linear fits.}
\label{Fig:fig3}
\end{figure}
\noindent\textbf{Thermoelastic strain relaxation in ML-WS$_2$ at 4\,K:\,}Estimating the local strain under cryogenic conditions is essential for analysing QD emission properties at 4\,K. To quantify strain relaxation upon cooling, we also measured 2D-X$^0$ (hereafter denoted as X$^0$) emission energies at multiple NP locations and in flat regions of ML-WS$_2$ at 4\,K for two samples S1 and S4. Figure\,\ref{Fig:fig2}b shows representative 4\,K $\mu$-PL spectra of ML-WS$_2$ acquired at the NP location (thin curve) and in the flat region (thick curve) of sample\,S1, marking X$^0$ energies with vertical lines corresponding to the respective locations. From measurements at multiple locations (see histograms in Supplementary Figs.\,S7g\textendash h), we evaluated the average temperature-induced X$^0$ energy shifts
\begin{center} $\Delta E_{avg}^{X0}\left(T = 296\rightarrow 4\,\text{K}\right)\,=\,E_{avg}^{X0}\left(4\,\text{K}\right)-E_{avg}^{X0}\left(296\,\text{K}\right)$,\end{center}
shown using column plots in Fig.\,\ref{Fig:fig2}c for both NP and flat regions. The measured $\Delta E_{avg}^{X0}\left(T = 296\rightarrow 4\,\text{K}\right)$ values exceed those expected from the Varshni relation for ML-WS$_2$\cite{nagler2018zeeman}, indicating an additional contribution due to the thermoelastic strain relaxation upon cooling to 4\,K. It likely arises from varying degrees of mechanical coupling between the ML-WS$_2$ flakes and the substrates or within the layered heterostructures.

We therefore quantify this strain relaxation by estimating the average local strain at NP locations at 4\,K using the same approach as at RT.  Figure\,\ref{Fig:fig2}d compares the average local strains of samples\,S1 and S4 at 4\,K (thin bars) and RT (thick bars), clearly indicating a significant reduction in tensile strain in both samples upon cooling. The strain relaxation, defined as $\Delta \epsilon_{\parallel}(T = 4\,\text{K}) = \epsilon_{\parallel}(4\,\text{K}) - \epsilon_{\parallel}(296\,\text{K})$, amounts to \textendash 0.36 $\pm$ 0.25\% (\textendash 0.21 $\pm$ 0.08\%) for samples\,S1 (S4), respectively, as shown in the inset of Fig.\,\ref{Fig:fig2}d. A smaller thermoelastic tensile strain relaxation of $\approx0.11$\% has previously been observed for thin GaAs membranes\cite{kumar2014anomalous} containing GaAs/AlGaAs QDs bonded to a substrate using the gold thermocompression technique. In contrast, the thermoelastic compressive strain relaxation in ML-WS$_2$ is significantly larger in magnitude; therefore, it is included in our subsequent analysis of QD emission at 4\,K. As the measured strain relaxation values for S1 and S4 overlap within their respective error bars, we adopt an average strain relaxation of \textendash 0.28\% and apply it to the RT average strains to estimate the corresponding 4\,K average strains for all samples.\\

\begin{figure}[htb]
\centering
\includegraphics[width=8.4cm]{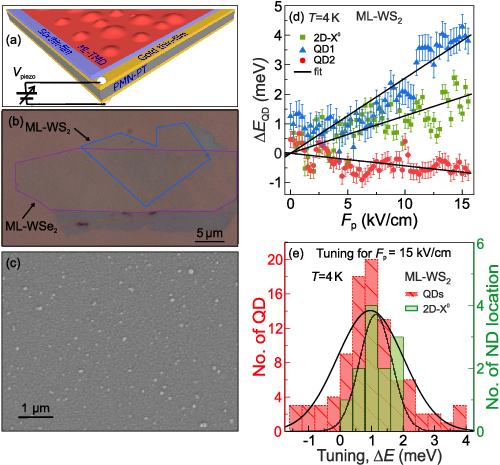}
\caption{\textbf{Statistical investigation of dynamic strain tuning of individual QDs and X$^0$ emission in ML-WS$_2$ at 4\,K.\,}(a) Schematic of ML-TMD flakes integrated onto a piezoelectric device D; layers are labeled. Strain in the ML-TMDs is induced by applying an electric field $\left(F_{\text{P}}\right)$ between the top and bottom surfaces of a piezoelectric PMN-PT substrate. (b) Optical micrograph of WS$_2$ and WSe$_2$ flakes transferred onto device D. ML regions of both materials are outlined by dotted lines. (c) SEM image of the SiO$_x$ surface of a piezo device, showing engineered SiO$_x$ nanodroplets (NDs) that locally strain the ML-TMDs. (d) $F_{\text{P}}$\textendash dependent emission energies of X$^0$(squares), QD1 (triangles), and QD2 (circles) from ML-WS$_2$ at ND locations. These are representative examples: the average tuning rate of X$^0$, and large positive and negative tuning rates of individual QDs, respectively. (e) Histogram of energy tuning at $F_{\text{P}}=15$\,kV\,cm$^{-1}$ for X$^0$ (textured bars) and QDs (closed bars) in ML-WS$_2$ at ND locations, highlighting the broad range of tuning observed for individual QDs (bin size: 0.4\,meV). Error bars in (d) are defined in the Methods section.}
\label{Fig:fig4}
\end{figure}
\noindent\textbf{Enhanced strain sensitivity of QD emission energy in ML-WS$_2$ and ML-WSe$_2$:\,}To investigate the role of local strain in tuning the QD ensemble emission, we plot the peak emission energies of QD ensembles in ML-WS$_2$ and ML-WSe$_2$ as a function of the average local strain at NP locations in multiple samples (Figs.\,\ref{Fig:fig3}a\textendash b). The observed redshift in the peak emission energies of the QD ensembles $\left(\Delta E_{\text{QD}}\right)$ across the samples, for both ML-WS$_2$ and ML-WSe$_2$, is consistent with an increase in tensile strain. The QDs exhibit linear trends (solid lines in Fig.\,\ref{Fig:fig3}a\textendash b) with gauge factors $\left(\Delta E_{\text{QD}}/\%\right)$ of -149\,$\pm$\,57\,meV/\% and -275\,$\pm$\,123\,meV/\% for ML-WS$_2$ and ML-WSe$_2$ samples, respectively. These values indicate large strain sensitivities: for QDs in ML-WS$_2$, $\Delta E_{\text{QD}}/\%$ is $ 3.9 \pm 1.5$ times larger than that of the X$^0$ gauge factor reported previously\cite{waheed2025large}, and for QDs in ML-WSe$_2$, it is $ 1.8 \pm 0.8$ times larger than the X$^0$ gauge factor reported previously\cite{roy2024upconversion}.

Earlier reports indicate a substantially lower $\Delta E_{\text{QD}}/\%$, in the range of 26\textendash 33\,meV/\%\cite{iff2019strain,chakraborty2020strain} for QDs in ML-WSe$_2$ when tuned using PMN-PT [100] substrates. In contrast, a significant enhancement of $\Delta E_{\text{QD}}/\%$\textemdash 120\,meV per \% strain was achieved using the PMN-PT [110] substrate\cite{iff2019strain}. For all samples discussed here, strain is created directly within ML-WSe$_2$ and ML-WS$_2$ flakes by NPs, thereby avoiding strain loss associated with transfer processes in PMN-PT platforms. We therefore attribute the observed large gauge factors to the direct creation of strain at the locations where the QDs are defined. Figure\,\ref{Fig:fig3}c compares gauge factors of QDs in TMD semiconductors with those of widely studied III\textendash V QDs (GaAs/AlGaAs\cite{kumar2011strain}, InGaAs/GaAs \cite{seidl2006effect,trotta2012nanomembrane}, InGaN/GaN\cite{holmes2021pure,jbara2016effect}), highlighting the enhanced strain sensitivity of QDs in ML-TMDs.\\

\begin{figure}[htb]
\centering
\includegraphics[width=8.5cm]{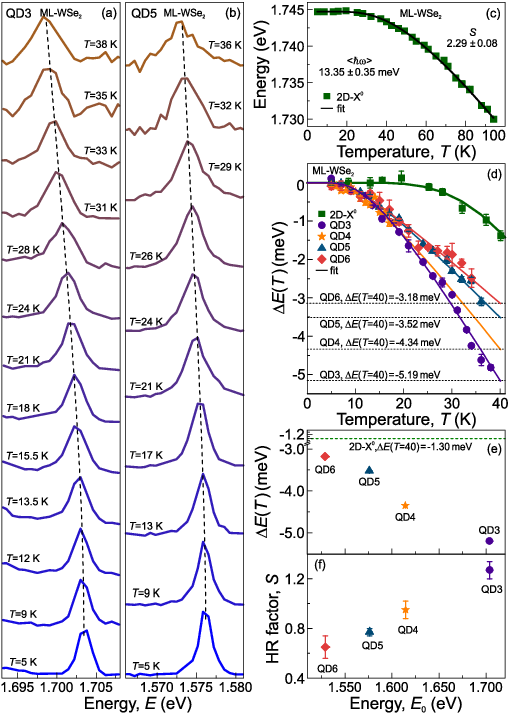}
\caption{\textbf{Interactions of phonons with QDs and X$^0$ in ML-WSe$_2$.\,}Temperature-dependent $\mu$-PL spectra of (a) QD3 and (b) QD5 in ML-WSe$_2$ from sample\,S6, illustrating the gradual redshift of their emission peaks as the temperature increases above 5\,K. (c) Temperature-dependent emission energies of X$^0$ and (d) corresponding changes in emission energies $\left(\Delta E(T)\right)$ of QD3 (circle), QD4 (star), QD5 (triangle), QD6 (diamond), and X$^0$ (square) in ML-WSe$_2$, demonstrating that QDs possess higher temperature sensitivity than X$^0$. The dotted lines in (a)\textendash (b) and solid lines in (c) \textendash (d) represent O'Donnell equation fits to $\Delta E(T)$. (e) $\Delta E(T)$ at $T$\,=\,40\,K and (f) Huang\textendash Rhys factors ($S$) of QDs as a function of their zero-temperature emission energies, together demonstrating enhanced electron-phonon interactions in QDs induced by quantum confinement. A dotted line in (e) represents $\Delta E(T)$ at $T$\,=\,40\,K for X$^0$.}
\label{Fig:fig5}
\end{figure}
\noindent\textbf{Dynamic strain tuning of QDs and X$^0$ in ML-WS$_2$ at 4\,K:\,}To further investigate the large strain sensitivities of QDs and the associated ensemble broadening (Fig.\,\ref{Fig:fig3}d\textendash e), we created dynamic strain in these ND-induced QDs in ML-WS$_2$ using piezoelectric device D (Fig.\,\ref{Fig:fig4}a). A wrinkle-free flake (optical micrograph in Fig.\,\ref{Fig:fig4}b) bends at NDs of size $\approx$75\,nm, creating QDs in a manner analogous to NP-induced QDs. SEM images of the NDs (Fig.\,\ref{Fig:fig4}b and Supplementary Fig.\,S14) were used to estimate their sizes. An electric field applied to the piezoelectric substrate (PMN\textendash PT) $\left(F_\text{P}\right)$ further modifies the static local strain, enabling dynamic control of strain in ML-WS$_2$ QDs.

We monitored $F_\text{P}$-dependent PL emission from QDs in ML-WS$_2$ at 25 different ND locations in device D. Each location showed several sharp emission lines, corresponding to the presence of multiple QDs. At an additional 12 ND locations, also showing QD emission, we measured the $F_\text{P}$-dependent X$^0$ emission alongside QD emission to compare the X$^0$ energy shifts $\left(\Delta E_{\text{X0}}\right)$ with those of the QDs $\left(\Delta E_{\text{QD}}\right)$. Figure\,\ref{Fig:fig4}d shows representative examples of the $F_\text{P}$ dependent energy shifts for QD1 (closed triangles), QD2 (closed circles), and X$^0$ (closed squares) measured at three ND locations. Histograms in Fig.\,\ref{Fig:fig4}e summarise the energy shifts ($\Delta E$) at $F_\text{P}$\,=\,15\,kV/cm, showing 93 QDs from 37 ND locations (textured bars) and X$^0$ from 12 ND locations (solid bars). A positive $F_\text{P}$ applied to the PMN-PT substrate of device D creates an in-plane biaxial compressive strain in the ML-WS$_2$ flake. Consequently, blueshifts in both X$^0$ and most QDs are observed, which are consistent with previous strain-tuning reports\cite{iff2019strain}. The small redshifts observed for 13 QDs, as shown in Fig.\,\ref{Fig:fig4}e, can arise due to the presence of QDs at apexes of the bends made by NDs\cite{iff2019strain}.\\

\noindent\textbf{Linewidth broadening of QD ensembles in ML-WS$_2$ and ML-WS$_2$ due to localised strain:\,}Notably, the coexistence of QDs with opposite energy shifts, together with variations in their magnitudes, leads to a broader distribution of $\Delta E_{\text{QD}}$ compared to that of $\Delta E_{\text{X0}}$ (Fig.\,\ref{Fig:fig4}e). These distributions are used to estimate the expected broadening of the QD ensemble under the tensile strain, as discussed below. The maximum $\Delta E_{\text{X0}}$ of 2.0\,meV corresponds to a tensile strain of 0.05\%. For this strain, we observe a total ensemble broadening of 5.40\,meV; comprising contributions of 3.80\,meV from the maximum blueshift and 1.60\,meV from the maximum redshift, as both shifts contribute to the overall QD energy distribution. Thus, we estimate a broadening of 108\,meV per \% biaxial strain for the QD ensemble in ML-WS$_2$.

Since we have measured the gauge-factors of QDs in both ML-WS$_2$ (see Fig.\,\ref{Fig:fig3}a) and ML-WSe$_2$ (see Fig.\,\ref{Fig:fig3}b), we can also estimate the ensemble broadening of QDs in ML-WSe$_2$ as follows. For QDs in ML-WS$_2$, a 1.0\,meV X$^0$ energy shift results in a 2.70\,meV broadening of the QD ensemble. The gauge factors of QDs in ML-WSe$_2$ and ML-WS$_2$ are 1.80 times and 3.90 times, respectively, relative to their corresponding X$^0$ gauge factors. Therefore, a 1.0\,meV X$^0$ energy shift (equivalent to $\approx$0.0065\% biaxial strain) in ML-WSe$_2$ is expected to produce a QD ensemble broadening of 1.25\,meV. Consequently, we obtain a strain-induced ensemble broadening of 192\,meV per \% biaxial strain for QDs in ML-WSe$_2$. Using the ensemble broadening rates of 108\,meV per \% strain for ML-WS$_2$ and 192\,meV per \% strain for ML-WSe$_2$, we performed a linear fit of the strain-dependent ensemble broadenings for all samples (Fig.\,\ref{Fig:fig3}d\textendash e). The fit gives intercepts $\omega_0$ of $ 67.5\,\pm\,5.0$\,meV for ML-WS$_2$ and $ 53.0\,\pm\,3.5$\,meV for ML-WSe$_2$. Strikingly, these $\omega_0$ values are in good agreement with the FWHM linewidths of X$^0$ emission peaks of 51.5\,$\pm$\,17.0\,meV for unstrained ML-WS$_2$ and 39.5\,$\pm$\,21.5\,meV for unstrained ML-WSe$_2$ (Supplementary Figs.\,S13c\textendash d).

The histogram of $\Delta E$ for QDs in ML-WS$_2$ (see Fig.\,\ref{Fig:fig4}e) also shows that $\approx$86\% QDs blueshift with a weighted-average energy-shift $\left(\Sigma\Delta E_{i}w_{i}/\Sigma w_{i}\right)$ of 1.26\,meV. However, X$^0$ emissions at all ND locations blueshift with a weighted-average energy-shift of 1.13\,meV. Although the piezo-tuning has created a small strain, a higher $\Sigma\Delta E_{i}w_{i}/\Sigma w_{i}$ of QDs compared to X$^0$ provides a possible explanation for the higher strain sensitivity observed for QDs in ML-WS$_2$. The literature\cite{Li04CdSe} suggests that the deformation potential (i.e., bandgap energy shift per unit volumetric strain) of a CdSe QD is higher than the bulk material, and its value increases as the quantum confinement is increasing due to its size reduction. Similar trends are expected for QDs in ML-TMDs due to quantum confinement effects, which we demonstrate in the following by performing temperature-dependent PL measurements.\\

\noindent\textbf{Interactions of phonons with QDs and X$^0$: }As the thermal expansion of a ML-TMD with increasing temperature\cite{Xuan_TMDexp18, Zhong_TMDexp22, Theresa_WSe2expan24} can be interpreted as thermally induced tensile strain, we performed temperature-dependent $\mu$-PL measurements on four single QDs in ML-WSe$_2$ to verify the role of quantum confinement in enhancing the strain sensitivity (or deformation potentials) of QDs in ML-TMDs. Figure\,\ref{Fig:fig5} summarizes these measurements and the extracted physical parameters of the QDs and of the X$^0$ from sample\,S6.

The evolution of the $\mu$-PL spectra for two QDs, QD3 and QD5, over the temperature range 5\textendash 38\,K is shown in Fig.\,\ref{Fig:fig5}a\textendash b, respectively (see Supplementary Fig.\,S16a\textendash b for $\mu$-PL spectra of the other two QDs, QD4 and QD6). At each temperature, the emission energies of the QDs and of the X$^0$ were extracted by fitting the spectral peaks with Gaussian line shapes. The $\mu$-PL spectra of the QDs were not recorded above $T$\,=\,38\,K because their emission rates became too low. In contrast, the $\mu$-PL spectra of the X$^0$ from a flat region of ML-WSe$_2$ were measurable up to 94\,K, the maximum temperature allowed by the cryostat's temperature controller.

The temperature dependence of the X$^0$ emission energy in semiconductors, including ML-TMDs, can be explained via the exciton-phonon coupling mechanism. This dependence is well described by the O'Donnell\textendash Chen equation,\cite{ODonnelChen91},\,$E(T)$\,=\,$E_0$\,-\,$S<\hbar\omega>\left[coth\left(<\hbar\omega>/2k_\text{B}T\right)-1\right]$, where $E_0$ is the zero-temperature emission energy, $S$ is the Huang-Rhys factor, $<\hbar\omega>$ is the average phonon energy, and $k_\text{B}$ is the Boltzmann Constant. As shown in Fig.\,\ref{Fig:fig5}c, the measured $E(T)$ values (solid squares) for the X$^0$ agree well with the O'Donnell\textendash Chen model (solid line), yielding $S$\,=\,2.29\,$\pm$\,0.08 and $<\hbar\omega>$\,=\,13.35\,$\pm$\,0.36\,meV. These values are consistent with previously reported parameters for ML-WSe$_2$\cite{arora2015excitonic,huang2016probing}.

The measured $E(T)$ values for all four QDs also agree well with the O'Donnell\textendash Chen model (Supplementary Fig.\,S16c\textendash f, and Tab.\,S2). Figure\,\ref{Fig:fig5}d shows the temperature-dependent energy shifts, $\Delta E(T)=E(T)-E_{0}$, of QDs and X$^0$ up to $T=40$\,K, where solid symbols and solid lines represent the measured and fitted values, respectively. The data clearly shows that for a 40\,K increase in temperature, the estimated redshifts for all four QDs (horizontal dashed lines) are significantly larger than that of the X$^0$, indicating stronger exciton-phonon coupling in the QDs compared to the delocalized X$^0$.\\
Interesting trends emerge when the energy shifts $\Delta E(T)$ and the Huang-Rhys factor $S$ are plotted against $E_{0}$ in Fig.\,\ref{Fig:fig5}e\textendash f, respectively. Both the redshift $\Delta E(T)$ and the value of $S$ increase with increasing $E_{0}$. As higher $E_{0}$ values signify stronger quantum confinement (i.e., lower confinement volume $V$), we attribute the enhanced redshifts and larger $S$ to quantum-confinement-driven strengthening of exciton-phonon coupling in the QDs. This interpretation is consistent with previous reports on semiconductor quantum-confined systems, where a scaling $S\propto 1/V$ has been proposed\cite{Schmitt87QD,Nestoklon22PbS}. Thus, these temperature dependent measurements validate our assumption that the higher strain sensitivities of QDs in ML-TMDs are driven by the quantum-confinement.
\section*{Discussion}
\noindent In conclusion, we demonstrate highly red-shifted QD emissions in ML-WS$_2$, with an energy shift of approximately 292\,meV, induced at an engineered nanoparticle location. Correlative $\mu$-PL and $\mu$-Raman spectroscopy reveal average in-plane local strain of up to 1\% at RT in ML-WS$_2$ (that relaxes by 0.28\% upon cooling to 4\,K) and 0.2\% at 4\,K in ML-WSe$_2$. By statistically analyzing thousands of QDs not only in ML-WS$_2$ but also in widely explored ML-WSe$_2$ material across multiple heterostructure samples and a strain-tuning piezoelectric device, we find that QDs in ML-TMDs exhibit strain sensitivities far exceeding those of conventional III-V QDs, reaching –149\,meV/\% for WS$_2$ and –275\,meV/\% for WSe$_2$. These values are significantly larger than their corresponding X$^0$ gauge factors by approximately 3.9 times for WS$_2$ and 1.8 times for WSe$_2$. These results establish strain-engineered TMD QDs as an exceptionally sensitive platform for tunable quantum emitters.

The dynamic strain induced by a piezoelectric device, combined with ND-generated static strain, produces large and diverse energy shifts per \% biaxial strain in QDs. Statistical analysis of energy shifts from a large ensemble of QDs and X$^0$ at QD locations in ML-WS$_2$ reveals that the observed ensemble linewidth broadenings of QDs in ML-TMDs\textemdash 108 meV per \% strain for ML-WS$_2$ and 192 meV per \% strain for ML-WSe$_2$\textemdash are driven by their enhanced strain sensitivities. Consistent with this interpretation, $\approx$\,86\% of the investigated QDs in ML-WS$_2$ show stronger blueshifts than X$^0$ under the in-plane compressive strain.

Temperature-dependent $\mu$-PL measurements further show that QD emission exhibits substantially larger redshift than X$^0$ emission for the same temperature increase between 5 and 40\,K, indicating stronger exciton–phonon coupling. Both the temperature-dependent energy shift $\Delta E(T)$ and Huang-Rhys factor $S$ increase with the zero-temperature emission energy $E_0$, demonstrating a clear correlation with stronger quantum confinement. These results indicate that confinement enhances the deformation potential, providing a microscopic origin for the enhanced strain sensitivities of QDs in ML-TMDs. Our findings establish a versatile strain-engineering strategy for spectral matching in solid-state, atomic, and hybrid quantum photonic networks, while offering new insights into phonon interactions in atomically thin QDs.

\section*{Methods}\label{sec11}
\noindent\textbf{Sample preparation:}
The ML-TMDs and hBN flakes were exfoliated from commercially available bulk crystals with the help of Nitto tape, and all-dry viscoelastic transfer techniques were utilised. Polydimethylsiloxane (PDMS) was used to transfer these flakes onto the required substrates. Thermal annealing at 150\textdegree C for 2 hours following each transfer was performed to eliminate PDMS residue from the top of the layer, thereby improving conformality for the subsequent layer and reducing moisture trapping. The dielectric (SiO$_2$) NPs (Sigma-Aldrich) with diameters ranging from 125-175\,nm were uniformly distributed on the required substrates by spin-coating an ethanol-based SiO$_2$ NP solution at 8000 rpm. SEM imaging was used to validate the particles' locations before transferring the ML-TMD flakes. We used SNPs, ENPs, NDs, and unintentional wrinkles as strain inducers.

ENPs were fabricated by depositing a 75 nm SiO$_2$ film after spin-coating the NPs using an e-beam physical vapour deposition (e-beam PVD) system. This additional SiO$_2$ film partially envelops each NP, enabling the monolayer to conform to the ENPs. In addition, the film enhances lateral structural support for the flake, significantly minimizing the potential for mechanical failure or rupture. Samples S1–S6 employ SNPs as strain inducers, whereas samples C1 and C2 use ENPs to generate localized strain. In samples W1 and W2, strain pockets in ML-WSe$_2$ arise from unintentional wrinkles formed during fabrication.

For dynamic tuning, an ND-induced piezoelectric device D was fabricated as follows. A Ti (5\,nm)/Au (100\,nm) film was first deposited on both sides of a 200\,$\mu$m-thick PMN-PT substrate for electrical contact. In order to eliminate plasmonic effects on the ML-TMDs, a thin SiO$_2$ film (110\,nm) was deposited on top of the Ti/Au film at a rate of 3.2\,$\pm$\,1.5\,\AA/s. All deposition processes were carried out in an e-beam PVD system under high vacuum. The high deposition rate of the SiO$_2$ film leads to the formation of NDs on the surface, which act as local strain inducers for ML-TMDs in device D.

Together, these ten samples fall under four distinct layered structures, as displayed in Fig.\,\ref{Fig:fig1}f. Samples\,S2, S3, S5, and S6 consist of hBN-encapsulated ML-WS$_2$ or ML-WSe$_2$, while sample\,S4 features ML-WS$_2$ isolated from the substrate by a bottom hBN layer. Sample\,C2 is identical to sample\,C1; in both samples, ML-WS$_2$ flakes were directly transferred onto substrates containing ENPs. Overall, QDs in ML-WS$_2$ were investigated in six samples (S1–S4 and C1–C2), while QDs in ML-WSe$_2$ were studied in four samples\textemdash S5–S6 with SNP-induced strain, W1–W2 with wrinkle-induced strain\textemdash as well as in device D with ND-induced strain under two different applied electric fields.

The device undergoes compression (tension) when a positive (negative) electric field is applied, and this strain is effectively transferred to the ML-TMDs attached to it. ML-WS$_2$ was first transferred onto NDs, and 37 distinct strained locations were investigated. Subsequently, ML-WSe$_2$ was transferred such that it partially covered the ML-WS$_2$. In this configuration, multiple locations exclusively within the ML-WSe$_2$ region were examined under varying applied fields to study statistical properties of QDs in ML-WSe$_2$.

\noindent\textbf{Optical characterization:}
$\mu$-Raman measurements were performed using a home-built confocal microscopy setup equipped with an objective with NA\,=\,0.75. $\mu$-PL measurements were also performed using a second home-built confocal microscopy setup equipped with an objective with NA\,=\,0.82. PL for the samples were first investigated at RT and subsequently at LT. For LT-PL measurements, the samples were mounted on an XY scanner integrated with a $\pm$2.5 mm XYZ nanopositioning stack (Attocube) inside a closed-cycle cryogen-free cryostat (AttoDry800), maintained at 4\,K. A diode-pumped solid-state continuous-wave (CW) laser with an excitation wavelength of 532 nm was used for both PL and Raman measurements.
An ultra-steep longpass filter with a cutoff wavelength of 533.3 nm (Semrock) was employed to block the laser and prevent its transmission into the spectrometer. All spectra were recorded using a 0.5\,m focal-length spectrometer equipped with a water-cooled charge-coupled device (CCD) camera, achieving a spectral resolution of approximately 125\,$\mu$eV (2.5 meV) at $\lambda$= 532\,nm with an 1800 lines/mm (150 lines/mm) grating. Temperature-dependent measurements were carried out using a heater (maximum power 5\,W) integrated with the piezo stack.

\noindent\textbf{Error estimation:}
The strain errors ($\Delta\epsilon_\text{PL}$) shown in Fig.\,\ref{Fig:fig2}a,d, where strains measured using PL measurements, are calculated using:
\begin{equation}
	\Delta\epsilon_\text{PL} = \epsilon_\text{PL}
	\times \sqrt{
		\left( \frac{\Delta E_{\text{err}}}{\Delta E} \right)^{2}
		+ \left( \frac{G_{\text{err}}}{G} \right)^{2}
	},
\label{strnerr}
\end{equation}
where $\epsilon_\text{PL}$ is the corresponding calculated strain, $\Delta E$ is the strain-induced shift in the 2D-X$^{0}$ emission energy with respect to a reference energy, and $\Delta E_\text{err}$ is the statistical error. The error bars correspond to $\pm\Delta\epsilon_\text{PL}$. The error propagation was applied when subtracting energy values. G and G$_{\text{err}}$ represent the gauge factor and its systematic error, respectively, both obtained from the literature.

Since Fig.\,\ref{Fig:fig2}c plots energy shift of average peak energies between RT and 4\,K, the error bars here are calculated using error of subtraction method using:
\begin{equation}
	\delta \Delta E_\text{err}\,=\,\sqrt{\left(\delta E_\text{err}^\text{RT}\right)^{2}
		+ \left(\delta E_\text{err}^\text{4\,K}\right)^{2}}
\end{equation}
where $\delta E_\text{err}$ is one-half standard deviation ($0.5\sigma$) spread of measurements for each energy. Figures\,\ref{Fig:fig3}a-b contain both X-axis and Y-axis errors. The strain error on the X-axis is calculated using Eq.\,\ref{strnerr}, while the energy error on the Y-axis is $0.5\sigma$ spread of measurements. Fits in Fig. 3a–b were performed considering both X- and Y-axes uncertainties.
The errors shown in Fig.\,\ref{Fig:fig3}c are the fitting errors with 1$\sigma$ confidences for data shown in Fig.\,\ref{Fig:fig3}a-b. In Figs.\,\ref{Fig:fig3}d-e, the strain error is determined using Eq.\,\ref{strnerr}, and the FWHM error corresponds to the fitting uncertainty of 0.5$\sigma$ confidence. In Fig.\,\ref{Fig:fig4}d, errors were estimated from the energy jittering measurements, and they are $0.5\sigma$. For Figs.\,\ref{Fig:fig5}c-d, $0.5\sigma$ spread of repeated measurements at a single temperature. Figure\,\ref{Fig:fig5}f use 1-$\sigma$ confidence fitting error as an error bar.
\section*{Data availability}
The data supporting this study are included in the Article and its Supplementary Information.
\backmatter

\bmhead{Acknowledgements}
We thank E. S. Kannan, S. R. Parne, and A. Rahman for the fruitful discussion and A. Rastelli for the data analysis software. This work was supported by the DST Nano Mission grant (DST/NM/TUE/QM-2/2019) and
the matching grant from IIT Goa. The authors also thank CoE-PCI, IIT Goa for access to Raman facility. I.D.P. thanks The Council of Scientific and Industrial Research (CSIR), New Delhi, for the doctoral fellowship. K.W. and T.T. acknowledge support from the JSPS KAKENHI (Grant Numbers 21H05233 and 23H02052) and World Premier International Research Center Initiative (WPI), MEXT, Japan.

\bmhead{Author contributions}
S.S. and Y.W. carried out $\mu$-PL and $\mu$-Raman measurements, supported by J.T.S. and I.D.P., under the supervision of S.K. S.S. and J.T.S. fabricated the samples under the supervision of S.K. K.W. and T.T. prepared the hBN material. S.S. and S.K. analyzed the data with support from Y.W. and I.D.P. S.K. led the interpretation of the results. S.S., Y.W., and S.K. wrote the manuscript. All authors discussed the results and contributed to the manuscript. S.K. conceived and coordinated the project.

\bmhead{Declarations}
Competing interests: The authors declare no competing interests.
\bmhead{Supplementary information}
Supplementary Figs. S1\textendash 16, Sections I\textendash IX and Tables S1\textendash 3.

\bibliography{references}% common bib file

\end{document}